\documentclass[aps,twocolumn]{revtex4}
\usepackage{amsmath,amsfonts,amssymb,graphics,graphicx,epsfig,color,bbm,texdraw}
\usepackage{subfigure}
\usepackage{fancyhdr,makeidx}
\newcommand{\R}{\mathbbm{R}}

\newcommand{\tr}{{\rm Tr}\,}
\renewcommand{\det}{{\rm Det}\,}
\newcommand{\gr}[1]{\boldsymbol{#1}}
\newcommand{\be}{\begin{equation}}
\newcommand{\ee}{\end{equation}}
\newcommand{\bea}{\begin{eqnarray}}
\newcommand{\eea}{\end{eqnarray}}
\newcommand{\ket}[1]{|#1\rangle}

\newcommand{\sig}{\gr{\sigma}}

\newcommand{\eq}[1]{Eq.~(\ref{#1})}

\newcommand{\eg}{\emph{e.g.}~}
\newcommand{\ie}{\emph{i.e.}~}

\begin{document}

\title{Detecting entanglement by symplectic uncertainty relations}
%Submitted for the Optical Quantum Information Science Special Issue

\author{Alessio Serafini}
%% for REVTeX4, each author name can be set in a separate \author{} field

\address{Institute for Mathematical Sciences, 
Imperial College London, London SW7 2PG, United Kingdom,\\
QOLS, Blackett Laboratory, Imperial College London, 
London SW7 2BW, United Kingdom, and\\
Department of Physics \& Astronomy, University College London, 
Gower Street, London WC1E 6BT, United Kingdom}

\email{serale@imperial.ac.uk}

\begin{abstract}
A hierarchy of multimode uncertainty relations on the second moments of $n$ pairs of 
canonical operators is derived in terms of quantities invariant under linear canonical 
({\ie}~`symplectic') transformations. 
Conditions for the separability of multimode continuous variable states
are derived from the uncertainty relations, generalizing 
the inequalities obtained in [Phys.~Rev.~Lett.~{\bf 96}, 110402 (2006)] to 
states with some transposed symplectic eigenvalues equal to $1$. 
Finally, to illustrate the methodology proposed for the detection of continuous variable entanglement, 
the separability of multimode noisy GHZ-like states is analysed in detail
with the presented techniques, 
deriving a necessary and sufficient condition for the separability of such states under an `even' bipartition of 
the modes. 
\end{abstract}

%\ocis{270.0270,270.2500}% REPLACE WITH CORRECT OCIS CODES FOR YOUR ARTICLE
                          % NOTE: \ocis{} IS ALIASED TO \pacs{} BUT MUST
                          % FORMAT THE TERMS CORRECTLY FOR EACH JOURNAL

\maketitle %% NULL FUNCTION WITH LATEX 2e; required for REVTeX4

\section{Introduction}
%Refer to \texttt{OSAstyle.tex} for guidelines on manuscript preparation, use of \mbox{Bib\TeX}, and similar. Our new style file (July 2003 version of \texttt{osajnl.sty}) has reduced line space to help save paper.
%To facilitate conversion, place all math in a proper math environment. For example, expression $3\times 4 = 12$ should be set this way, \texttt{\$3$\backslash$times 4=12\$}, not this way, \texttt{3 \$$\backslash$times\$4=12}.

The effective detection of entanglement in continuous variable systems 
has become a task of crucial importance, notably in view of the rising relevance 
of such systems for quantum information and 
communication protocols, especially in quantum optical settings \cite{review}.
To this aim, several methods have been proposed lately, variously emphasising 
theoretical and practical aspects of the matter. 

In first seminal contributions to this aim,
the positivity of the partial transposition has been shown to be 
necessary and sufficient for the 
separability of $(1+n)$-mode Gaussian states ({\ie}~of states with Gaussian Wigner function 
shared by two subsystems of $1$ and $n$ modes respectively) \cite{simon00,werewolf}.
Furthermore, a general -- `computationally efficient', but not analytical -- criterion has been found 
for the separability of any Gaussian state under any bipartition, 
based on the iteration of a nonlinear map \cite{giedkeprl01}.
More recently, a full set of conditions has been developed to detect genuine multipartite 
entanglement in multimode states \cite{vanloock03} and a complete theory 
of optimised linear and `curved' entanglement witnesses 
for second moments (capable of detecting even bound entangled states) has been developed \cite{hyllus06}.
Also, a hierarchy of inequalities for higher moments (generally sufficient for entanglement) 
has been obtained \cite{vogel05}, allowing to significantly improve the entanglement qualification 
of non Gaussian states as well.

This paper focuses on the separability 
of multimode continuous variable states under any bipartition.
The approach of Ref.~\cite{simpleprl} , introducing a class of (generally) 
sufficient conditions for entanglement 
on second statistical moments, will be recovered and discussed in detail.
Moreover, such an approach is extended to a more general class 
of inequalities, capable of dealing with `pathological' instances 
(entangled states which were not detected by the original inequalities derived in Ref.~\cite{simpleprl}).
Finally, a relevant class of multimode continuous variable states (``GHZ-like'')
will be explicitly considered and analysed with the presented methodology, yielding 
a compact necessary and sufficient condition for the separability of such states 
of $2n$-mode systems under $(n+n)$-mode bipartitions.
It is worth remarking that, 
as is clearly the case for all conditions on second moments alone, the experimental 
test of the inequalities we will derive does not require a complete tomography 
of the state under examination -- generally a daunting task -- but only the 
measurement of the second moments, which might turn out to be delicate for multimode systems 
but seems in the reach of present techniques \cite{laurat}.

The inequalities we will present are based on a class of uncertainty relations
on second moments,
which may be compactly expressed in terms of symplectic invariants 
({\em i.e.~}of quantities invariant under linear canonical transformations).
Besides retaining a major fundamental interest {\em per s\`e}, uncertainty relations
have also lately been acknowledged as a main ingredient in the detection 
and qualification of entanglement, whose general scope goes well beyond 
the scenario of continuous variable systems \cite{guehneprl04,entropic,entropic2,sanders05}. 
In fact, for any state $\varrho$ of a bipartite quantum system, 
the positivity of the partially transposed density matrix $\tilde{\varrho}$ 
(obtained from $\varrho$ by transposing the Hilbert space of only one of the two 
subsystems) is a necessary condition for the state to be separable \cite{peres96,horodecki96}. 
In other words, the violation of the positivity of $\tilde{\varrho}$ is a proof of the presence 
of quantum entanglement in the state $\varrho$, 
which can in such a case be exploited to various quantum informational aims.
Now, uncertainty relations for quantum observables 
derive only from the commutation relations and from the positivity of the generic density matrix $\varrho$. 
Therefore, any relation 
derived from an uncertainty relation by replacing the state $\varrho$
with the partially transposed state $\tilde{\varrho}$ provides a way of testing the positivity 
of $\tilde{\varrho}$ and constitutes thus 
a sufficient condition for the state $\varrho$ to be entangled.
In the present instance, since the original uncertainty relations come in terms of 
second moments alone, the entanglement conditions are cast in terms of second moments as well,
and are therefore of great experimental relevance.
Furthermore, as it will be shown, 
these conditions turn out to be also necessary for the presence of 
entanglement whenever 
the positivity of the partial transpose is also
sufficient for separability, namely in $(1+n)$-mode Gaussian states and
$(m+n)$-mode bisymmetric Gaussian states \cite{werewolf,serafozzi05,boundnote}.

This paper is organised as follows. Canonical systems of many modes (like discrete bosonic fields in second quantization or 
motional degrees of freedom of material particles in first quantization), notation and uncertainty relations 
are introduced in Sec.~\ref{uncecan}. 
In Sec.~\ref{invasec}, quantities invariant under symplectic 
operations on the field modes are constructed as functions of the second moments of the field operators.
In terms of such invariant quantities, simple uncertainty relations for the second moments of 
any $n$-mode system are derived in Sec.~\ref{uncesym}. 
The partial transposition of such a relation will promptly lead to conditions for entanglement 
in terms of the second moments for any bipartition of the modes in Sec.~\ref{enta}. 
Finally, specific instances of multimode states are considered in Sec.~\ref{exempla} and conclusions 
are drawn in Sec.~\ref{conclu}.

\section{Uncertainty relations for canonical systems}\label{uncecan}

Let us consider a quantum mechanical system 
described by $n$ pairs of canonically conjugated 
operators $\{\hat x_j,\hat p_j\}$, 
each of them satisfying the canonical commutation relations (CCR).
As well known, by virtue of the Stone-von Neumann theorem the CCR admit, 
for any finite $n$, only 
a unique, infinite dimensional representation, thus allowing for the occurrence of
continuous spectra for the canonical operators. 
Therefore the system in case, whose variables could be motional degrees of freedom of 
particles in first quantization or quadratures of a bosonic field in second quantization, 
is commonly referred to as a ``continuous variable'' (CV) system. 
Grouping the canonical operators together in the 
vector $\hat R=(x_1,p_1,\ldots,x_n,p_n)^{\sf T}$ allows to 
concisely express the CCR \cite{ccrnote} (with $\hbar=2$ for future convenience) as
\be
[\hat R_j,\hat R_k] = 2i \Omega_{jk} \; , \label{ccr}
\ee
where the symplectic form $\Omega$ is defined by
\be
\Omega \equiv \oplus_{1}^{n} \omega \; , \quad \omega\equiv \left(\begin{array}{cc}
0 & 1 \\
-1 & 0 
\end{array}\right) \, .
\ee
A major role in the description of the system is played by the transformations 
which, acting in Heisenberg picture on the vector of operators $\hat R$, preserve the
fundamental commutation relations set out in \eq{ccr}. 
All the conceivable dynamical evolutions of the system have to comply with such a prescription. 
In particular, \eq{ccr} shows that allowed transformations acting {\em linearly} 
on $\hat R$ must preserve the symplectic form $\Omega$ under congruence: such 
linear transformations form the {\em real symplectic group} $Sp_{2n,\R}$ 
(reality being necessary to
preserve the hermiticity of the canonical operators):
$$
S\in SL({2n,\R}) \, :\;
S\in Sp_{2n,\R} \Leftrightarrow S^{\sf T} \Omega S = \Omega \; .
$$ 
At the Hilbert space level, these transformations correspond to unitary operations
generated by second order polynomials in the canonical operators. This fact, rigorously
justified by the construction of the infinite dimensional 
`metaplectic' representation of the symplectic group (allowed by the Stone-von Neumann theorem),
can be heuristically understood by considering the Heisenberg equation of motion for 
$\hat R$ under the action of a quadratic Hamiltonian and by recalling \eq{ccr}.

Clearly, any state of a $n$-mode CV system is described by a hermitian, positive, trace-class 
operator $\varrho$. Let us define the matrix of second moments, or ``covariance matrix'' (CM), 
${\gr\sigma}$ (with entries $\sigma_{jk}$) of the state $\varrho$ as
\be
{\sigma}_{jk} \equiv \tr{[(\hat R_j  \hat R_k +\hat R_k  \hat R_j) \varrho]}/2
-\tr{[\hat R_j \varrho]}\tr{[\hat R_k \varrho]} \; .
\ee
The CM is a symmetric $2n\times 2n$ matrix. The positivity of $\varrho$ 
and the CCR imply the following semidefinite 
constraint to be satisfied by any {\em bona fide} CM ${\gr\sigma}$ \cite{simon87,simon94}
\be
{\gr\sigma} + i\Omega \ge 0 \; . \label{obsheis}
\ee
This well known inequality (whose proof is, 
for the ease of the reader and to provide a self-contained exposition, 
reported in Appendix A)
is the only constraint a symmetric $2n\times2n$ matrix has to satisfy
to qualify as the CM of a physical state. 
For future convenience, let us write down the CM of an $n$-mode system in terms of 
$2\times 2$ submatrices ${\gr\gamma}_{jk}$
\be
{\gr\sigma} = \left(\begin{array}{ccc}
{\gr\gamma}_{11} &\cdots& {\gr\gamma}_{1n} \\
\vdots & \ddots & \vdots \\
{\gr\gamma}_{n1} &\cdots & {\gr\gamma}_{nn}
\end{array}\right) \, . \label{subba} 
\ee
Symmetry implies $\gr{\gamma}_{jk}=\gr{\gamma}_{kj}^{\sf T}$.
Let us also notice that, 
because of the skew-symmetry of $\Omega$, Inequality (\ref{obsheis}) ensures 
the strict positivity of ${\gr\sigma}$ \cite{posinote}:
\be
{\gr\sigma} > 0 \; . \label{posi}
\ee 

For a single mode system, the uncertainty principle (\ref{obsheis}) can be more explicitly 
recast as 
\be
\det{{\gr\sigma}}\ge 1 \; , \label{robsch}
\ee
which corresponds to the well known Robertson-Schr\"odinger uncertainty relation
\cite{robertson29,schrodinger30}. 
Inequality (\ref{robsch}) together with the condition (\ref{posi}) are {\em equivalent} 
to the uncertainty principle (\ref{obsheis}) for single-mode systems. 
For a two-mode system, the uncertainty principle reads 
\be
\det{{\gr\sigma}} + 1 \ge \Delta^{2}_{1} \; , \label{sera}
\ee
where $\Delta^{2}_{1}\equiv\sum_{j,k=1}^{2}\det{{\gr\gamma}_{jk}}$. 

Note that the quantities $\det{{\gr\sigma}}$ and $\Delta^{2}_{1}$, 
entering into Inequalities (\ref{robsch}) and (\ref{sera}), are
invariant under symplectic transformations \cite{serafozzi04}.
Clearly, one could provide a complete set of conditions determining the physicality of the CM 
by just verifying the positivity of the matrix $\sig+i\Omega$, which can be checked 
by considering the determinants of the principal submatrices (see, {\em e.g.}, \cite{hj}). 
Still, the invariant nature of the uncertainty principle implies that 
the explicit expression of the uncertainty relation for a general $n$-mode CM $\sig$
must be possible in terms of symplectic invariants constructed from 
the entries of ${\gr\sigma}$. 
In the following section we shall single out such invariants, 
by which an explicit, general 
expression for a multimode uncertainty relation will be subsequently derived.

\section{Construction of the symplectic invariants}\label{invasec}

To begin with let us recall that, because of the positivity condition (\ref{posi}),
one can apply a seminal result by Williamson \cite{williamson,williamson2,williamson3} (originally concerning the 
classification of quadratic Hamiltonians) to the positive quadratic form ${\gr\sigma}$ to infer 
the following result, which constitutes a basic tool of symplectic analysis:  
for any CM ${\gr\sigma}$ there exists a (non-unique) symplectic transformation $S\in Sp_{2n,\R}$
such that 
$$
S^{\sf T} {\gr\sigma} S = {\gr\nu} \; , 
$$
where 
$$
{\gr \nu} = \oplus_{j=1}^{n} \,{\rm diag}\,{(\nu_j,\nu_j)} \; . 
$$
The quantities $\{\nu_j\}$ are referred to as {\em symplectic eigenvalues}, 
while the transformation $S$ is said to perform a {\em symplectic diagonalisation} 
of the CM ${\gr\sigma}$. The matrix ${\gr\nu}$ is the so called Williamson normal form of 
the CM ${\gr\sigma}$. 
The uncertainty principle (\ref{obsheis}) can be succintly recast in terms 
of the symplectic eigenvalues $\{\nu_j\}$ as 
\be
\nu_j\ge 1  \label{eigheis}
\ee
for $j=1,\ldots,n$. Inequality (\ref{eigheis}) is fully equivalent to the uncertainty relation 
(\ref{obsheis}). Its proof is immediate, since it derives from the straightforward 
computation of the eigenvalues of the matrix ${\gr\nu}+i\Omega$, where ${\gr\nu}$ is the 
normal form associated to the CM ${\gr\sigma}$.

We can proceed now to the construction of the symplectic invariants for an $n$-mode 
state. The symplectic eigenvalues are of course symplectic invariants but, for an $n$-mode
system, their analytical expression in terms of the second moments turns out to be 
rather cumbersome (when possible at all).
As a matter of fact, the symplectic eigenvalues can be computed by diagonalising 
the matrix $\Omega{\gr\sigma}$, whose eigenvalues turn out to be $\{\mp i\nu_j\}$ 
for $j=1,\ldots,n$. The latter statement is easily proved by checking it on the normal form 
${\gr \nu}$ and by considering that 
\be
\Omega{\gr\nu} = \Omega S^{\sf T}{\gr\sigma} S = S^{-1}\Omega{\gr\sigma} S  
\quad {\rm for\; some}\;S\in Sp_{2n,\R} \, . \label{dictat}
\ee
Acting by similarity, the transformation $S$ preserves 
the spectrum of $\Omega{\gr\sigma}$ which thus coincides with the one of $\Omega{\gr\nu}$.
Now, a natural choice of symplectic invariants, dictated by \eq{dictat}, is 
given by the principal minors of the matrix $\Omega{\gr\sigma}$, which are 
invariant under symplectic transformations acting by congruence on ${\gr\sigma}$.
As we will see shortly, a compact and elegant uncertainty principle 
can be expressed in terms of such invariants 
for any number of modes. 
Let $M_{k}(\alpha)$ be the 
principal minor of order $k$ of the matrix $\alpha$ \cite{minornote}. 
Then, we define the symplectic invariants of a $n$-mode state $\{\Delta^n_j\}$ 
for $j=1,\ldots,n$ as
\be
\Delta^n_j \equiv M_{2j}(\Omega{\gr\sigma}) \; . \label{definva}
\ee
The principal minors of odd order vanish because of the alternate sign in the 
spectrum of $\Omega{\gr\sigma}$, leaving us with $n$ independent 
symplectic invariants $\{\Delta^n_j\}$
(as one should have expected, since the number of symplectic eigenvalues is $n$).
The quantities $\{\Delta^{n}_j\}$ are also known as 
``quantum universal invariants'' \cite{dodonov}.
We are now interested in the expression of the invariants $\{\Delta^n_j\}$ in terms of 
the symplectic eigenvalues $\{\nu_j\}$, which can be straightforwardly retrieved by 
considering the normal form ${\gr\nu}$ and turns out to be the following
\be
\Delta^n_j = 
\sum_{{\cal S}^{n}_{j}} \prod_{k\in{\cal S}^{n}_{j}} \nu_{k}^{2} \; , \label{inva}
\ee
where the sum runs over all the possible $j$-subsets ${\cal S}^{n}_{j}$ 
of the first $n$ natural integers ({\em i.e.~}over all the possible combinations of 
$j$ integers smaller or equal than $n$). 
In the following, $\Delta^n_j(\nu_1,\ldots,\nu_n)$ will stand for the functions relating the 
symplectic eigenvalues to the symplectic invariants, according to \eq{inva}.
Clearly, one has $\Delta^n_n=\det{{\gr\sigma}}$ while, for two-mode states, it is immediate to verify
that the invariant $\Delta^2_1$ coincides with the quantity appearing 
in Inequality (\ref{sera}) (this correspondence is clarified and generalised
in Appendix B, where an alternative way to compute the invariants $\{\Delta^n_j\}$ is 
described).

The following section is devoted to understanding how the uncertainty relation 
constrains the values allowed for the symplectic invariants of a $n$-mode state, 
giving rise to `symplectic' uncertainty relations.

\section{Symplectic uncertainty relations}\label{uncesym}

Let us first introduce some further notation. 
Let us consider an $n$-mode CV system and define the quantity ${\Sigma}_n$ as
\be
{\Sigma}_n = \sum_{j=0}^{n} (-1)^{n+j} \Delta^n_j \; , \label{sig}
\ee
where we assume $\Delta^n_0 \equiv 1$. 
Likewise, ${\Sigma}_n(\nu_1,\ldots,\nu_n)$ will stand for the function relating the 
symplectic eigenvalues $\{\nu_j\}$ to ${\Sigma}_n$, according to Eqs.~(\ref{inva}) and
(\ref{sig}).\smallskip

\noindent {\bf Symplectic uncertainty relation:} {\em Let ${\gr\sigma}$ be the covariance matrix 
of a $n$-mode continuous variable state.
The symplectic invariant ${\Sigma}_n$, determined according to Eqs.~(\ref{definva}) and (\ref{sig}), 
fulfills the inequality 
\be
{\Sigma}_n \ge 0 \; .  \label{symheis}
\ee
}\medskip

\noindent {\em Proof.} The statement (obviously true for $n=1$, for which it reduces to
$\nu_1^2\ge1$) will be proven by induction. 
Let us then assume that, for a generic $n$, ${\Sigma}_{n-1}\ge0$.
One has $\partial_{\nu_{k}^2}{\Sigma}_n (\nu_1,\ldots,\nu_n) =
{\Sigma}_{n-1}(\nu_1,\ldots,\nu_{k-1},\nu_{k+1},\ldots,\nu_n)\ge 0$ 
(because of the inductive hypothesis) for any $k\le n$ and for any value of the $\{\nu_j\}$. 
Therefore, because of the bound (\ref{eigheis}),
the minimum of ${\Sigma}_n$ is attained for $\nu_{j}=1$ $\forall j\le n$. 
In such a case one has 
$$
{\Sigma}_n = (-1)^{n} \sum_{j=1}^{n} (-1)^{j} C^n_j = 0 \; ,
$$
where $C^n_j$ is the bynomial coefficient. 
Inequality (\ref{symheis}) is thus established. $\Box$ \smallskip

Alternately it may be shown, exploiting Eqs.~(\ref{inva}) and (\ref{sig}), that 
\be
{\Sigma}_n = \prod_{j=1}^{n} (\nu_j^2-1) \; . \label{veritas}
\ee

Inequality (\ref{symheis}), reducing to the well known relations (\ref{robsch}) and
(\ref{sera}) for, respectively, $n=1$ and $n=2$, 
provides a general and elegant
way of expressing a necessary uncertainty relation constraining the symplectic invariants. 
Rather remarkably,
such a relation can be analytically checked for any (finite) number of bosonic degrees of freedom.
However \eq{veritas} shows that, actually, Inequality (\ref{symheis}) is 
only necessary and not sufficient for the full uncertainty relation (\ref{eigheis})
to be satisfied as it is not able to detect unphysical CMs for which an even number of 
symplectic eigenvalues violates Inequality (\ref{eigheis}).

It is worth remarking that 
this impossibility is pertinent to the considered uncertainty relation and is not due to any 
fundamental lack of information in the symplectic invariants.
To better clarify this point, let us first consider the simple case $n=2$: 
in this instance Inequality (\ref{symheis}) cannot spot the unphysical case $\nu_j<1$ for $j=1,2$.
However, this situation can be ruled out by verifying the additional proviso 
$\det{\gr\sigma}\ge1$. Summarizing, the uncertainty principle (\ref{eigheis}) is {\em equivalent},
for two-mode states, to the set of conditions
\be
\Delta^2_2-\Delta^2_1+\Delta^2_0 \ge 0 \, , \quad
\Delta^2_2\ge 1 \; , \quad
{\gr\sigma}>0 \; . 
\ee 
(The case $\nu_j<0$ cannot be discriminated by symplectic invariants, 
since they are functions of the squared symplectic eigenvalues: the additional 
condition ${\gr\sigma}>0$ has thus always to be required.)

More generally, for $n$-mode systems, the knowledge of the symplectic invariants 
allows for retrieving the symplectic eigenvalues, so that the complete 
uncertainty relation (\ref{eigheis}) can be in principle recast in terms of symplectic invariants alone.
To show this fact, let $\nu$ denote any one of the symplectic eigenvalues, say $\nu_1$, and 
let $\Delta_{k}^{n-1}=\Delta_k^{n-1}(\nu_2,\ldots,\nu_n)$. Then one has 
\be
\Delta^n_k = \nu^2 \Delta_{k-1}^{n-1}+\Delta_k^{n-1} \quad {\rm for}\;\; k=1,\ldots,n \, , \label{eigsyst}
\ee
where $\Delta_{n}^{n-1}=0$ is understood. Once the symplectic eigenvalues $\{\Delta^n_j\}$ 
are known, the quadratic system (\ref{eigsyst}) for the unknowns $\nu$ and $\{\Delta_k^{n-1}\}$  
can be solved. 
Such a system of $n$ equations allows for $n$ sets of solutions, 
providing the $n$ symplectic eigenvalues as the $n$ solutions for $\nu$.  
This procedure provides an efficient way of deriving the symplectic eigenvalues 
of the CM ${\gr\sigma}$ and yields, in principle, a recipe to recast the full
uncertainty relation in terms of symplectic invariants. 

\eq{veritas} also allows to elucidate another particular facet of the uncertainty relations. 
Suppose that one symplectic eigenvalue $\nu_{j}$ satisfies $\nu_{j}=1$.
As apparent, in such a case  
the symplectic uncertainty relation (\ref{symheis}) is saturated. However,
the complete uncertainty relation (\ref{eigheis}) is not necessarily fully saturated, as the other symplectic 
eigenvalues could be greater than one. This situation is described as 
{\em partial saturation} of the uncertainty relation. 
`Partially saturating' states have been proven to 
be endowed with peculiar properties. For instance, concerning two-mode Gaussian states, 
partially saturating states turn out to coincide with the states with minimal entanglement for 
given purities \cite{adesso03,adesso032}, which in turn prove to be relevant in characterizing 
the multipartite entanglement structure of three-mode Gaussian states \cite{adesso06}.

Actually, even for states featuring partial saturation a simple necessary uncertainty relation 
may always be found in terms of the quantities $\{\Delta^{n}_j\}$. 
In fact, let us suppose that $p<n$ symplectic eigenvalues $\nu_j$ 
are equal to $1$. In this case, one can iteratively exploit \eq{eigsyst}, for $\nu=1$, 
to express all the invariants of lower order $\{{\Delta}^{n-p}_{j}\}$ for $j=0,\ldots,n-p$ 
in terms of the original invariants $\{{\Delta}^{n}_{j}\}$, obtaining 
\be
{\Delta}^{n-p}_{j} = (-1)^p \sum_{k_1=j+1}^{n-p+1}\sum_{k_2=k_1+1}^{n-p+2}\ldots
\sum_{k_p=k_{p-1}+1}^{n} (-1)^{j+k_{p}} {\Delta}^{n}_{k_{p}} \, , \label{itero}
\ee
where $\{k_j\}$ is a set of $p$ summation indexes. 
Finally, one can recast the uncertainty relation of lower order
$\sum_{j}(-1)^{n-p+j}{\Delta}^{n-p}_{j}\ge 0$ simply as 
\be
\begin{split}
\sum_{j=0}^{n-p}\sum_{k_1=j+1}^{n-p+1}\ldots
\sum_{k_p=k_{p-1}+1}^{n} (-1)^{n+k_{p}} {\Delta}^{n}_{k_{p}} &=\\
\sum_{j=p}^{n} \left[(-1)^{n+j} p\,! C^j_p \Delta_{j}^{n} \right]  &\ge 0 \, , \label{corrigheis}
\end{split}
\ee
where $C^j_p$ is the binomial coefficient defined as customary by $C^j_p=j\,!/[(j-p)\,!p\,!] $.
\eq{corrigheis} cannot be saturated if, as per hypothesis, only $p$ symplectic eigenvalues are equal to $1$. 
Therefore, in case of saturation of Inequality (\ref{symheis}), one may proceed to test Inequality (\ref{corrigheis})
for increasing values of $p$, until the saturation disappears: the first nonzero value will determine the 
legitimacy of the CM.
%Notice that Inequalities (\ref{itero}) and (\ref{corrigheis}) take very simple forms for small values of $p$. 
%For $p=1$, one has
%\be
%{\Delta}^{n-1}_{j} = \sum_{k=j+1}^{n} (-1)^{j+k+1} {\Delta}^{n}_{k} \, , \label{itero1}
%\ee
%leading to
Note that, in general, Inequality (\ref{corrigheis}) 
hold for partially saturating states, {\em when at least $p$ 
symplectic eigenvalues are equal to $1$} (in fact, the inequality has been derived under such an assumption).

In the next section, criteria for the separability and entanglement of quantum states 
will be obtained from the uncertainty relation previously derived.

\section{Symplectic separability criteria}\label{enta}

The positivity of the partially transposed state (``PPT criterion'') 
is a necessary condition for the separability of any bipartite quantum state
({\em i.e.~}for the possibility of creating the state by local operations 
and classical communication alone). 
Conversely, the violation of such positivity is a sufficient condition 
for a quantum state to be entangled, in which case 
quantum correlations are at disposal and may be exploited for quantum informational tasks.
Moreover, as far as the CV systems here addressed are concerned, 
the PPT criterion turns out to be sufficient as well for the separability of $(1+n)$-mode
Gaussian states ({\em i.e.~}of $(1+n)$-mode states with Gaussian Wigner and characteristic functions) 
and of bisymmetric $(m+n)$-mode Gaussian states
(here and in what follows, we refer to a bipartite `$(m+n)$-mode' CV state as to a state
separated into a subsystem $A$ of $m$ modes, owned by party $A$, and a  
subsystem $B$ of $n$ modes, owned by party $B$).

These facts come in especially handy for CV systems, as the action of partial transposition 
on covariance matrices is easily described. Let $\varrho$ be a $(m+n)$-mode bipartite CV state 
with $2(m+n)$-dimensional CM ${\gr\sigma}$. 
Then the CM $\tilde{{\gr\sigma}}$ of 
the partially transposed state $\tilde{\varrho}$ with respect to, say, subsystem $A$, 
is obtained by switching the signs of the $m$ momenta $\{p_{j}\}$ belonging to subsystem $A$.
In formulae:
\be
\tilde{{\gr\sigma}} = T {\gr\sigma} T \, , \quad {\rm with} \quad 
T \equiv \oplus_{1}^{m} \left(\begin{array} {cc}
1&0\\
0&-1
\end{array}\right) \oplus {\mathbbm 1}_{2n} \, , \label{transpo}
\ee
where ${\mathbbm 1}_{2n}$ stands for the $2n$-dimensional identity matrix.
Now, in analogy with Inequality (\ref{obsheis}) 
derived from the positivity of the density matrix $\varrho$, 
a (generally) sufficient condition for separability derived by the PPT 
criterion is given by \cite{simon00}
\be
\tilde{{\gr\sigma}}+i\Omega \ge 0 \;  \label{obssepa}
\ee
or, in terms of the symplectic eigenvalues $\{\tilde{\nu}_j\}$ 
of the partially transposed CM $\tilde{{\gr\sigma}}$
(whose normal form will be henceforth denoted by $\tilde{\gr\nu}$), as
\be
\tilde\nu_j\ge 1 \; . \label{eigsepa}
\ee
The results of the previous section allow one to recast such separability criteria
for $(m+n)$-mode states in terms of partially transposed symplectic invariants 
$\{\tilde{\Delta}^{m+n}_{j}\}$, defined by $\tilde{\Delta}^{m+n}_{j}\equiv 
M_{2j}(\Omega{\tilde{\gr\sigma}})$.

This simple accessory result will be precious in what follows: \smallskip

\noindent{\bf Little lemma.}
{\em Let ${\gr\sigma}$ be the physical CM of a state of a $(m+n)$-mode CV system,
with $m\le n$. 
Let $\tilde{{\gr\sigma}}$ 
be the partial transposition of ${\gr\sigma}$ with respect to any of the two subsystems. 
Then, at most $m$ symplectic eigenvalues $\tilde{\nu}_j$ of $\tilde{\gr\sigma}$ can violate
Inequality (\ref{eigsepa}).}\medskip

\noindent{\em Proof.} Suppose to perform the transposition in the $m$-mode subsystem:
$\tilde{\gr\sigma}=T_{m}{\gr\sigma}T_{m}$, where $T_m$ is the partial transposition 
matrix in the $m$-mode subsystem (equal to the matrix $T$ of \eq{transpo}).
Let ${\cal D}(\alpha)$ be the dimension of the subspace upon which 
the generic matrix $\alpha$ is negative definite. 
Since $T_m$ reduces to the identity on a $(2n+m)$-dimensional subspace, Inequality (\ref{obssepa})
reduces to the (definitely satisfied) Inequality (\ref{obsheis}) on such a subspace, thus implying 
${\cal D}(\tilde{{\gr\sigma}}+i\Omega)\le m$. 
One has then ${\cal D}(\tilde{\gr\nu}+i\Omega)={\cal D}(\tilde{{\gr\sigma}}+i\Omega)\le m$,
where the equality holds because the signature is preserved under congruence transformations 
(`Sylvester's inertia law') and 
$\tilde{\gr\nu}+i\Omega=S^{T}(\tilde{{\gr\sigma}}+i\Omega)S$ 
for some $S\in Sp_{2(m+n),\R}$. Straightforward computation shows that the eigenvalues of 
$\tilde{\gr\nu}+i\Omega$ are given by $\{\tilde{\nu}_j\mp 1\}$, thus proving the result 
as the $\{\tilde\nu_j\}$ have to be positive ($\tilde{\gr\sigma}=T_m{\gr\sigma} T_m>0$ because
${\gr\sigma}>0$). 
The choice of the transposed subsystem is not relevant, 
since $(T_n T_m) (\tilde{\gr\sigma}+i\Omega) (T_n T_m)=\tilde{\gr\sigma}'-i\Omega$
(where $T_n$ and $\tilde{\gr\sigma}'$ are, respectively, 
the partial transposition matrix and the partially transposed CM with respect to the $n$-mode subsystem), 
and $\tilde{\gr\sigma}'-i\Omega\ge0$ is equivalent to $\tilde{\gr\sigma}'+i\Omega\ge0$. $\Box$\smallskip

In analogy with \eq{sig}, let us now define
\be
\tilde{{\Sigma}}_{m+n} = \sum_{j=0}^{m+n} (-1)^{m+n+j} \tilde\Delta^{m+n}_j \; . \label{sig2}
\ee
The inequality 
\be
\tilde{{\Sigma}}_{m+n} \ge 0 \; , \label{symsepa}
\ee
being necessary for Inequality (\ref{eigsepa}) to be satisfied, 
is a necessary condition for separability under $(m + n)$-mode bipartitions 
and is thus a sufficient condition to detect 
entanglement in such a multimode system, irrespective of the nature of the state under examination.

Let us remark that, while for Gaussian states Inequality (\ref{eigsepa}) (which ultimately 
reduces to a condition on the second moments) is necessary and sufficient for the positivity of the 
partial transpose, this is not the case for non-Gaussian states.
For such states conditions resorting to higher moments lead to a sharper 
detection of negative partial transposition, and thus of entanglement \cite{vogel05}.

Even for Gaussian states, Inequality (\ref{eigsepa}) cannot detect the negativity 
of the partial transpose 
if an even number of symplectic eigenvalues violate condition (\ref{eigsepa}).
However, let us focus on Gaussian states under $(1 + n)$-mode bipartitions 
(for which the PPT criterion is necessary and sufficient for separability \cite{werewolf}). 
Because of the previous lemma, for these states 
at most one partially transposed symplectic eigenvalue can violate Inequality (\ref{eigsepa}). 
Inequality (\ref{symsepa}) is then capable of detecting such a violation.

The same argument applies to `bisymmetric' Gaussian states, 
defined as the $(m+n)$-mode Gaussian states which are invariant under 
mode permutations internal to the $m$-mode and $n$-mode subsystems. 
A bisymmetric Gaussian state with CM $\gr\sigma$ can be reduced, by local symplectic operations 
(on the $m$-mode and $n$-mode subsytems), 
to the tensor product of a two-mode Gaussian state and of 
uncorrelated thermal states \cite{serafozzi05}, with global CM $\gr\sigma_{2}$:
$\gr\sigma_2 = S_l^{\sf T} \gr\sigma S_l$ for some $S_l\in\,Sp_{2m,\R}\oplus Sp_{2n,\R}$. 
The lemma above 
can be applied to obtain ${\cal D}(T\gr\sigma T+i\Omega)\le1$
from which, observing that
$$
(TS_l^{\sf T}T) (T\gr\sigma T+i\Omega)(TS_lT)  = 
T\gr\sigma_2 T+i\Omega \; ,
$$
one infers that at most one 
partially transposed symplectic eigenvalue of the CM $\gr\sigma$ can violate Inequality (\ref{eigsepa}).
Notice that the locality of the operation $S_l$ is crucial in establishing this result, since it implies 
$(TS_l^{\sf T}T)\Omega (TS_lT)=\Omega$ (which would not generally hold for a $S_l$ with nondiagonal 
terms relating the $m$-mode to the $n$-mode subsystem). 
 
Inequality (\ref{symsepa}) is thus necessary and sufficient for the separability 
of all $(1+n)$-mode and bisymmetric $(m+n)$-mode Gaussian states,
except for the set of `null measure' for which any
$\tilde{\nu}_j$ is identical to $1$ (that could be entangled but have $\tilde{\Sigma}_{m+n} = 0$).
However, as in the case of the uncertainty relation,
a necessary and sufficient condition for separability 
in terms of the quantities $\{\tilde{\Delta}_j\}$ may be found for such states as well. 
If $p$ symplectic eigenvalues $\tilde{\nu}_j$ (for $p\le n$) 
are equal to $1$, such a relation reads, in analogy with Inequality (\ref{corrigheis}),
\be
\sum_{j=p}^{m+n} \left[(-1)^{m+n+j} p\,! C^j_p 
\tilde{\Delta}_{j}^{m+n} \right]  \ge 0 \; . \label{corrige}
\ee
\eq{corrige} cannot be saturated if, as per hypothesis, only $p$ eigenvalues were equal to $1$. 
Therefore, in case of saturation of Inequality (\ref{symsepa}), one may proceed to test Inequality (\ref{corrige})
for increasing values of $p$, until the saturation disappears: the first nonzero value will determine 
the positivity of the partial transposition of the state at the level of second moments.
Notice that the case for which $\tilde{\nu}_j=1$ for some $j$ is indeed very relevant in 
designing and detecting entangled resources for quantum information as, 
typically, it corresponds to {\em pure} CV states.
For $p=0$, Inequality (\ref{corrige}) correctly reduces to Inequality (\ref{symsepa}), 
which has thus been generalised.
Summing up, {\em Inequality (\ref{symsepa})} -- together with Inequalities (\ref{corrige}) 
in case of partial saturation --  
{\em are necessary and sufficient for the separability 
of $(1+n)$-mode and of bisymmetric $(m+n)$-mode Gaussian states.}

The simple condition (\ref{symsepa}) on the second moments,
which can be promptly analytically verified, 
may be very helpful in the detection of interesting CV entangled states 
(in the same spirit as in Refs.~\cite{vanloock03,hyllus06}). 
This aspect is 
especially relevant in view of the recent experimental developments 
in the implementation of multipartite CV quantum information protocols,
usually relying on symmetric resources (discriminated by the previous condition) \cite{yonezawa04}.

The next section will better illustrate 
the efficacy of Inequality (\ref{symsepa}) for the detection of CV entanglement,
by addressing the separability of a relevant example of multimode states.

\section{Example: noisy GHZ-like states}\label{exempla}

The CV ``GHZ-type'' states\cite{namenote} are a class of fully symmetric multimode 
Gaussian states introduced in Ref.~\cite{vanloock00} as the prototypical resource for the 
implementation of a CV teleportation network. 
Experimentally, they can be generated 
by inserting squeezed vacua into an array of beam splitters 
(for a detailed description of the generating scheme, see \cite{vanloock03} and \cite{adesso06bis}).
Moreover, they turn out to be 
the symmetric Gaussian states maximising both the couplewise (between any pair of modes) 
and the genuine multipartite entanglement \cite{adesso05,adesso06} and their multipartite 
entaglement proves to be especially resilient to decoherence \cite{adesso06}.
We will study here the separability of GHZ-type states determining the partially transposed 
symplectic invariants and employing Inequality (\ref{symsepa}).

To fix ideas, we will consider a GHZ-type 
state with an even number of total modes $2n$ under 
the $n+n$ bipartition. 
The CM $\sig$ of such a state, generated from impinging squeezed vacua
with squeezing parameter $r$, is described by \eq{subba} with $2\times 2$ blocks
given by 
\bea
\gr{\gamma}_{jj} = q \left(\begin{array}{cc} 
\frac{{\rm e}^{2r}+ (2n-1){\rm e}^{-2r}}{2n} & 0 \\
0 & \frac{(2n-1){\rm e}^{2r} + {\rm e}^{-2r}}{2n}
\end{array}\right) \, ,&
\quad 1\le j \le 2n  \, , \nonumber\\
&\\
\gr{\gamma}_{jk} = q \left(\begin{array}{cc} 
\frac{\sinh(2r)}{n} & 0 \\
0 & -\frac{\sinh(2r)}{n}
\end{array}\right) \, ,& 
\quad \; j\neq k \; . \nonumber
\eea
Realistically, we have assumed a thermal noise with mean photon number $(q-1)\ge 0$ 
to affect the creation of the states (this amounts to multiplying the CM by $q$). 

After some basic algebra (see also Appendix B), the partially transposed minors $\tilde\Delta^{n+n}_j$ can be 
explicitly worked out and found to be
\be
\tilde{\Delta}^{n+n}_{j} = q^{2j}\left[(C^{2n}_{j}-g^{n}_j) + g^{n}_j \cosh(4r)\right] \; ,
\ee
where the coefficients $g^{n}_j$ are determined by the following sums
\be
g^{n}_j \equiv \left(\sum_{k=0}^{j-1} (-1)^{j+k-1} 2^{j-k} C^{2n}_{k} \right) - 
\left( \sum_{k=0}^{j-3} (-1)^{j+k-1} 2^{j-k-2} C^{2n}_k \right)\; 
\ee
and satisfy a recursive relation which will shortly be useful
\be
g^{n}_j = 2(C^{2n}_{j-1}-g^{n}_{j-1})-g^{n}_{j-2} \quad {\rm for}\; 2\le j \le 2n \, . \label{recur}
\ee
Also, one has $g^{n}_0 = 0$ and $g^{n}_1=2$. 
Inequality (\ref{symsepa}) then reads
\be
\sum_{j=0}^{2n} (-1)^{j+2n} q^{2j}\left[(C^{2n}_{j}-g^{n}_j) + g^{n}_j \cosh(4r)\right] 
\label{ghzsepa}\ge 0 \, .
\ee
Now, the polynomial in $q$ in the LHS of Inequality (\ref{ghzsepa}) admits $4n-4$ degenerate roots 
for $q=\mp1$ and $4$ more roots for $q=\mp\,{\rm e}^{\mp2r}$ [as can be shown by employing 
\eq{recur}]. Because the polynomial is obviously diverging for infinite $q$ (the leading order in $q$ has 
a positive sign) and accounting for the physical condition $q\ge 1$, we find that Inequality (\ref{ghzsepa})
is equivalent to 
\be
q \ge \,{\rm e}^{2r} \; . \label{ghzsepagen}
\ee
Such a simple criterion, derived from Inequality (\ref{symsepa}), 
is necessary and sufficient for the separability of $2n$-mode noisy GHZ-like states 
under $(n+n)$-mode bipartitions, regardless of the total number of modes
(let us recall that, in the absence of noise, 
such states are already known to be always inseparable under any bipartition). 
Note also that the physical significance of the derived condition is immediately evident: 
in order to maintain the entanglement in the final state,
the thermal noise has to be counterbalanced with a corresponding level of squeezing in the
initially uncorrelated input modes.

Notably, an analogous analysis -- based on Inequality (\ref{symsepa}) --
may be applied for any number of modes under any bipartition 
and for more general states, 
getting more and more useful as the number of modes increases.

\section{Discussion and outlook}\label{conclu}

The separability criteria here presented, compactly cast in terms of symplectic invariants, 
have been proven to be necessary and sufficient for two relevant classes of Gaussian states and
have led to remarkably simple conditions for the separability of GHZ-like states. 
Besides this immediate usefulness,
the presented analysis, deriving from a recasting of the 
uncertainty relations in terms of symplectic invariants,
touches the very core of the symplectic structure, 
encompassing the full description of Gaussian states of light 
and discrete bosonic systems in general.
More specifically, it further clarified the specific constraints imposed by quantum mechanics 
on the second moments of canonical operators, in a naturally 
canonically invariant framework. 

This fundamental analysis might yield further interesting results 
concerning the entanglement characterization of CV states. 
In particular, 
the parametrisation of Gaussian states through symplectic invariants 
has provided remarkable insight into the entanglement properties of two-mode
states \cite{adesso03,adesso032} and 
could be, employing the techniques here presented, 
carried over to the analysis of multipartite continuous variable entanglement,
which has been lately drawing considerable attention 
\cite{giedke01,wolf04,adesso04,adesso05,adesso06,adessomarie,hiroshima,zhang05}
mostly in view of the remarkable experimental perspectives uprising
in quantum optical systems.

\subsection*{Acknowledgments}
Helpful discussions with D.~Gross and K.~Audenaert are acknowledged.
This work has been carried out as part of a Marie Curie Intra-European Fellowship.

%After the manuscript is proofread, the {\tt .tex} file and figures
%should be archived with tar-gzip compression. Do not include
%subdirectories within the archive.

%To upload your manuscript, follow the instructions on the each
%journal's homepage (see \mbox{\href{http://www.opticsinfobase.org}{http://www.opticsinfobase.org}}).
%Authors should feel free to contact OSA staff for assistance; details are available at \href{InfoBase}{http://www.opticsinfobase.org}.

%\appendix

%\section*{Appendix A: Sample}
%\setcounter{equation}{0}
%\renewcommand{\theequation}{A{\arabic{equation}}}

%\begin{equation}
%a+b=c.
%\end{equation}

\appendix

\section*{Appendix A: Proof of Inequality (\ref{obsheis})}\label{appe0}
\setcounter{equation}{0}
\renewcommand{\theequation}{A{\arabic{equation}}}

Let us consider the operator $\hat y = (\hat R-\,{\rm Tr}\,[\varrho\hat R])^{\sf T}Y$, 
where $Y\in{\mathbbm C}^{2n}$ is an arbitrary complex vector. 
The positivity of $\varrho$ implies ${\rm Tr}\,[\varrho\hat{y}^{\dag}\hat y]\ge 0$,
from which 
\be
{\rm Tr}\,[\varrho\hat{y}^{\dag}\hat y] = 
Y^{\dag}\gr\tau Y = Y^{\dag}(\gr\sigma+i\Omega)Y \ge 0 \, , \label{equi}
\ee
where the matrix $\gr\tau$ has entries $\tau_{jk}\equiv\,{\rm Tr}\,[\varrho\hat R_j \hat R_k]
-\,{\rm Tr}\,[\varrho R_j]\,{\rm Tr}\,[\varrho R_k]$ and 
the CCR (\ref{ccr}) have been employed. 
Due to the arbitrarity of $Y\in {\mathbbm C}^{2n}$, Inequality (\ref{equi}) corresponds 
to Inequality (\ref{obsheis}), which is thus proven. 

Notice how the uncertainty relation has been derived 
assuming solely the CCR (\ref{ccr}) and the positivity of the density matrix $\varrho$.

\section*{Appendix B: Computation of the symplectic invariants}\label{appe}
\setcounter{equation}{0}
\renewcommand{\theequation}{B{\arabic{equation}}}
An alternative way of computing the symplectic invariants 
defined by \eq{inva} is here outlined, providing some deeper insight  
into the symplectic architecture of covariances of bosonic systems.

Let ${\bar m}=\{m_1,\ldots,m_k\}$ stand for a ordered $k$-subset of natural integers
smaller or equal than $n$, 
such that $m_j\in{\mathbbm N}$ and $n\ge m_j>m_{j-1}$ for $j=1,\ldots,k$, 
and let ${\cal N}^{k}_{n}$ be the set of all such $k$-subsets.
Now, for $\bar l, \bar m \in {\cal N}^k_n$, let us define the $2k\times2k$ submatrix 
${\gr\sigma}^k_{{\bar l},{\bar m}}$ of the CM ${\gr\sigma}$ as 
\be
{\gr\sigma}^k_{{\bar l},{\bar m}} = \left(\begin{array}{ccc}
{\gr\gamma}_{l_1 m_1} &\cdots& {\gr\gamma}_{l_1 m_k} \\
\vdots & \ddots & \vdots \\
{\gr\gamma}_{l_k m_1} &\cdots & {\gr\gamma}_{l_k m_k}
\end{array}\right) \; ,
\ee
where the $2\times 2$ submatrices ${\gr\gamma}_{jk}$ are defined as in \eq{subba}. 

The symplectic invariants $\Delta^{n}_{k}$ are then given by
\be
\Delta_{k}^{n} = \sum_{{\bar l},{\bar m}\in {\cal N}^{k}_n} \det{{\gr\sigma}^k_{{\bar l},{\bar m}}} \; ,
\ee
where the sum runs over all the possible ordered $k$-subset. 
This definition amounts to considering the sums of the determinants of all the 
possible (also non principal) $2k\times 2k$ submatrices of ${\gr\sigma}$ obtained by 
selecting all the combinations of $2\times 2$ blocks (each block describing 
one mode or the correlations between a pair of modes).
For instance, for any $n$, one has $\Delta_1^n=\sum_{j,k=1}^{n}\det{{\gr\gamma}_{jk}}$,
that is just the sum of the determinants of the $2\times 2$ blocks themselves 
(this is the multimode generalisation of the invariant entering in the 
uncertainty relation (\ref{sera}) for two-mode states).

%\newpage

%\section*{List of Figure Captions}

%Fig. 1. Multipanel figure assembled into one file with proper
%arrangement and labeling.
%\noindent Fig. 2. ...

%\noindent Fig. 3. ...

%\newpage
%% sample figure environment
%  \begin{figure}[htbp]
%  \centering
%  \includegraphics[width=8.3cm]{OT10000F1.eps}
%  \caption{Multipanel figure assembled into one EPS file with proper arrangement and labeling. AO10000F1.eps.}
  %% \label{}
%  \end{figure}


\begin{thebibliography}{99}
\bibitem{ccrnote} The CCR could be, more rigorously, introduced in terms 
of the (bounded) displacement operators $D_{\xi}=\exp{(i\hat R^{\sf T}\Omega\xi)}$, 
for $\xi\in\R^{2n}$, as $D_\xi^{\dag}D_{\xi'}=D_{\xi'}D_{_\xi}^{\dag}\exp{(4i\xi^{\sf T}\Omega\xi')}$.
However, to our aims, the simpler, straightforward definition given by Eq.~(\ref{ccr}) will suffice.
Note that, through all the paper, it is understood that the second moments of the field operators
$\{\hat R_j\}$ are finite for the considered states.
\bibitem{namenote}{Such states draw their name by their expression in the quadrature basis 
in the limit of infinite squeezing that reads, up to normalisation, 
$\int \ket{x,\ldots,x} \,{\rm d}x$, where $\ket{x,\ldots,x}$ 
stands for the simultaneous eigenvector of the $n$ position operators with equal eigenvalue $x$.}
\end{thebibliography}
\end{document}